\documentstyle[12pt,epsf]{article}
\begin{document}
\renewcommand{\thetable}{\Roman{table}}
\def \beq{\begin{equation}}
\def \cd{c^+ \dn}
\def \cu{c^+ \up}
\def \dd{d^+ \dn}
\def \du{d^+ \up}
\def \dn{\!\!\downarrow}
\def \Dn{\!\!\Downarrow}
\def \eeq{\end{equation}}
\def \G{{\rm GeV}}
\def \M{{\rm MeV}}
\def \s{(2)^{-1/2}}
\def \st{(3)^{-1/2}}
\def \ss{(6)^{-1/2}}
\def \sd{s^+ \dn}
\def \su{s^+ \up}
\def \ud{u^+ \dn}
\def \uu{u^+ \up}
\def \up{\!\!\uparrow}
\def \Up{\!\!\Uparrow}
\def \vac{|0 \rangle}
\rightline{DOE/ER/40561-218-INT95-17-05}
\rightline{EFI-95-48}
\rightline{hep-ph/9508252}
\vspace{0.5in}
\centerline{\bf CHARMED BARYONS WITH $J = 3/2$
\footnote{To be submitted to Phys.~Rev.~D.}}
\vspace{0.5in}
\centerline{\it Jonathan L. Rosner}
\centerline{\it Institute for Nuclear Theory}
\centerline{\it University of Washington, Seattle, WA 98195}
\bigskip
\centerline{and}
\bigskip
\centerline{\it Enrico Fermi Institute and Department of Physics}
\centerline{\it University of Chicago, Chicago, IL 60637
\footnote{Permanent address.}}
\bigskip

\centerline{\bf ABSTRACT}
\medskip
\begin{quote}
The width of a recently discovered excited charmed-strange baryon, a candidate
for a state $\Xi_c^*$ with spin 3/2, is calculated. In the absence of
configuration mixing between the ground-state (spin-1/2) charmed-strange baryon
$\Xi_c^{(a)}$ and the spin-1/2 state $\Xi_c^{(s)}$ lying about 95 MeV above it,
one finds $\tilde \Gamma(\Xi^*_c \to \Xi_c^{(a)} \pi) = (3/4) \tilde
\Gamma(\Xi^* \to \Xi \pi)$ and $\tilde \Gamma(\Xi^*_c \to \Xi_c^{(s)} \pi) =
(1/4) \tilde \Gamma(\Xi^* \to \Xi \pi)$, where the tilde denotes the partial
width with kinematic factors removed.  Assuming a kinematic factor for P-wave
decay of $p_{\rm cm}^3$, one predicts $\Gamma(\Xi^*_c \to \Xi_c^{(a)} \pi) =
2.3$ MeV, while the $\Xi^*_c \to \Xi_c^{(s)} \pi$ channel is closed.  Some
suggestions are given for detecting the $\Sigma_c^*$, the spin-3/2 charmed
nonstrange baryon, and the $\Omega_c^*$, the spin-3/2 charmed doubly-strange
baryon.
\end{quote}
\newpage

\centerline{\bf II.  INTRODUCTION}
\bigskip

Candidates for all the ground-state baryons with a single charmed quark and
total spin equal to 1/2 have now been observed.  These consist of the
isosinglet $\Lambda_c(2285) = udc$, isotriplet $\Sigma_c(2453) =
(uuc,~udc,~ddc)$, and isodoublet $\Xi_c(2468) = usc,~dsc$ states listed by the
Particle Data Group \cite{PDG}, the isosinglet $\Omega_c(2704) = css$
\cite{omc}, and an excited $\Xi_c$ lying about 95 MeV/$c^2$ above the lowest
$\Xi_c$ and decaying to it by photon emission \cite{xics}.\footnote{The
numbers in parentheses denote the masses in MeV/$c^2$.}  These states are
depicted as the solid lines in Fig.~1. However, until recently the only
candidate for a spin-3/2 state was a cluster of six events produced by
neutrinos in a heavy-liquid bubble chamber \cite{Ammosov}, corresponding to a
$\Sigma_c^*$ state at $2530 \pm 5 \pm 5$ MeV/$c^2$ not yet confirmed in other
experiments.

\begin{figure}
\centerline{\epsfysize = 4in \epsffile{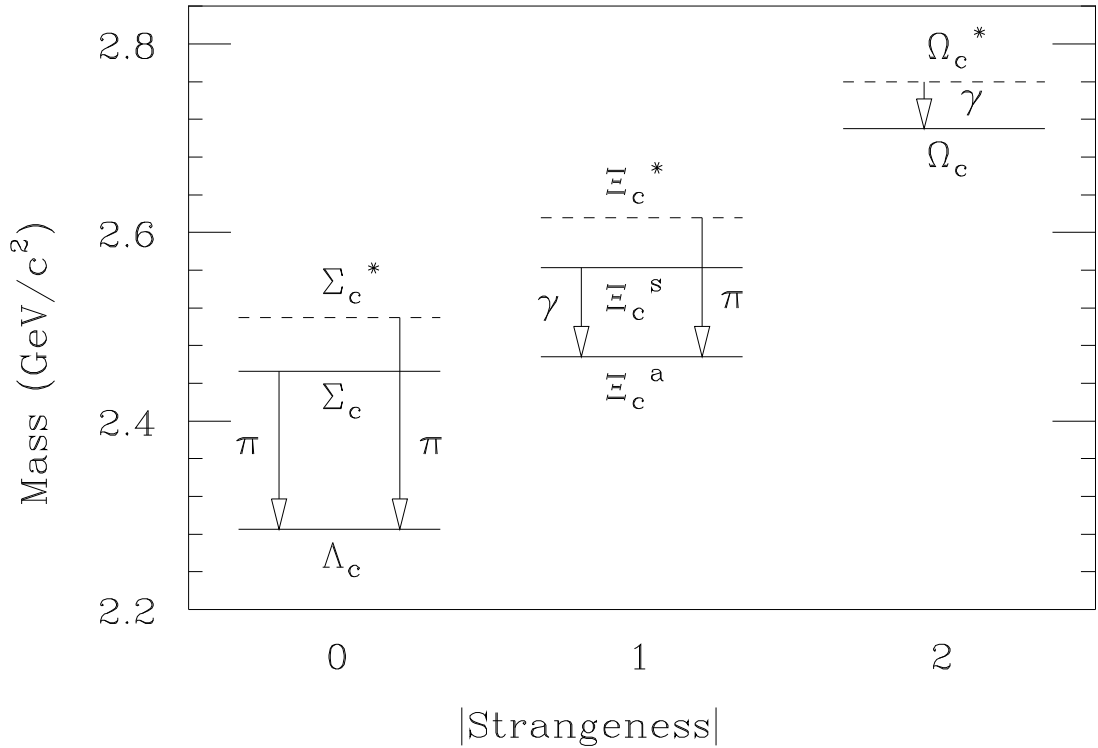}}
\caption{Lowest-lying S-wave states of a single charmed quark and two light
($u,~d,~s$) quarks.  Solid and dashed lines correspond to states with total
quark spin equal to 1/2 and 3/2, respectively.  Masses of the spin-1/2 states
and the $\Sigma_c^*$ and $\Xi_c^*$ correspond to observed values (see text),
while mass of the $\Omega_c^*$ is the lower limit predicted in the present
work. Superscripts on the spin-1/2 $\Xi_c$ states denote antisymmetry (a) and
symmetry (s) with respect to interchange of light-quark spins.  Transitions
are denoted by arrows.}
\end{figure}

The CLEO Collaboration has now presented evidence \cite{xicstar} for a narrow
state decaying into $\Xi_c^+ \pi^-$.  The state lies $178.2 \pm 0.5 \pm 1.0$
MeV/$c^2$ above the $\Xi_c^+$.  It has a width less than 5.5 MeV (90\% c.l.),
and has been identified by the authors as a candidate for the $\Xi_c^*$ shown
by the dashed line in Fig.~1, a spin-3/2 charmed baryon.

In the present article we predict the $\Xi_c^*$ to have a total width of 2.3
MeV (or less, if symmetry breaking effects are important), narrower than the
experimental resolution in the CLEO experiment.  This small width may be the
reason for the prominence of the signal.  By contrast, the $\Sigma_c^*$ is
expected to have a larger total width as a result of a larger matrix element
and a larger phase space for its $\Lambda_c \pi$ decay mode.  Nonetheless, its
predicted partial width is not expected to be so large that it should be
unobservable. We shall argue that the $\Sigma_c^*$ should be no heavier than
2552 MeV/$c^2$ and its total width should not exceed 35 MeV.  We also predict
the mass of the $\Omega_c^*$ to be at least 2771 MeV/$c^2$.

We introduce notation and calculate decay matrix elements in Section II.  The
decays of non-charmed baryons are discussed in Sec.~III, while charmed baryon
decays are described in Sec.~IV.  We conclude in Sec.~V.
\bigskip

\centerline{\bf II.  NOTATION AND CALCULATION OF MATRIX ELEMENTS}
\bigskip

\leftline{\bf A.  Representation of baryon states}
\bigskip

We describe pion emission using a quark model language \cite{RWu} which sums
the amplitudes for transitions of individual $u$ or $d$ quarks. We represent
baryon states by the action of three bosonic quark creation operators on the
vacuum, thereby taking account of antisymmetry with respect to color. We denote
the operator which creates a quark $q$ with $J_z = +1/2$ by $q^+ \up$.  A
baryonic state with $J_z = +1/2$ is denoted by $|B \up \rangle$, while one with
$J_z = 3/2$ will be denoted by $|B \Up \rangle$.  Thus, for example, a
$\Delta^{++}$ with $J_z = 3/2$ would be written $| \Delta^{++} \Up \rangle =
\ss(\uu \uu \uu) \vac$, where the factor is the usual one ($n!^{-1/2}$)
associated with $n$ identical Bose particles.  The spin-lowering operator $S_-$
may then be used to construct $| \Delta^{++} \up \rangle = \s (\uu \uu \ud)
\vac$.  The isospin-lowering operator $I_-$ gives us $| \Delta^+ \up \rangle =
\ss (\uu \uu \dd +~2 \uu \du \ud) \vac$.  We may then construct the proton with
$J_z = 1/2$ as the state orthogonal to this: $|p \up \rangle = \st (\uu \uu \dd
-~\uu \du \ud) \vac$.

The quark model states needed for the present calculations are given in Table
I.  Other states may be obtained by applying isospin raising or lowering
operators.  We shall need only states with $J_z = 1/2$ since we will be
concerned only with emission of (spinless) pions, so we denote the states $|B
\up \rangle$ merely by $B$ in the Table.  The full set of non-charmed states
has been given in Ref.~\cite{RWu}, whose sign conventions we adopt here.

The charmed ($C = 1$) states may be obtained from the non-charmed ($C = 0$)
ones by simple substitutions.  For example, the $\Lambda_c = udc$ is obtained
from the $\Lambda = uds$ by the replacement $s \to c$.  A similar replacement
converts a $\Sigma^+ = uus$ to a $\Sigma_c^{++} = uuc$. The state $\Xi_c^{+(a)}
= usc$, in which the $u$ and $s$ quarks are coupled to $J = 0$, is obtained
from the $\Lambda = uds$ by the replacements $d \to s$, $s \to c$.  Similarly,
the state $\Xi_c^{+(s)}$, in which the $u$ and $s$ quarks are coupled to $J =
1$, is obtained from the $\Sigma^0$ by the same replacements. The hyperfine
interaction between the light quarks is attractive in the $\Xi_c^{(a)}$ and
repulsive in the $\Xi_c^{(s)}$, leading to $M(\Xi_c^{(a)}) < M(\Xi_c^{(s)})$.
We shall ignore configuration mixing \cite{mix} between the $\Xi_c^{(a)}$ and
$\Xi_c^{(s)}$ states.  Similar types of substitutions may be applied to the $J
= 3/2$ states. For example, we obtain $\Xi_c^{*+}$ from $\Sigma^{*0}$ by
replacing $d \to s,~s \to c$.

\renewcommand{\arraystretch}{1.3}
\begin{table}
\caption{Quark model baryon states with $J_z = 1/2$ in terms of bosonic
creation operators acting upon the vacuum.}
\begin{center}
\begin{tabular}{c c c} \hline
Multiplet & State & Configuration \\ \hline
$J = 1/2$ & $p$             & $\st(\uu \uu \dd -~\uu \ud \du) \vac$ \\
($C = 0$) & $\Lambda$       & $\s(\uu \dd \su -~\ud \du \su) \vac$ \\
          & $\Sigma^+$      & $\st(\uu \ud \su -~\uu \uu \sd) \vac$ \\
          & $\Sigma^0$      & $\ss(\uu \dd \su +~\ud \du \su -~2 \uu \du \sd)
\vac$ \\
          & $\Xi^0$         & $\st(\su \su \ud -~\su \sd \uu) \vac$ \\ \hline
$J = 1/2$ & $\Lambda_c^+$   & $\s(\uu \dd \cu -~\ud \du \cu) \vac$ \\
($C = 1$) & $\Sigma_c^{++}$ & $\st(\uu \ud \cu -~\uu \uu \cd) \vac$ \\
          & $\Xi_c^{+(a)}$  & $\s(\uu \sd \cu -~\ud \su \cu) \vac$ \\
          & $\Xi_c^{+(s)}$  & $\ss(\uu \sd \cu +~\ud \su \cu -~2\uu \su \cd)
\vac$ \\ \hline
$J = 3/2$ & $\Delta^{++}$   & $\st(\uu \uu \ud) \vac$               \\
($C = 0$) & $\Sigma^{*+}$   & $\ss(\uu \uu \sd +~2 \uu \ud \su) \vac$ \\
          & $\Sigma^{*0}$   & $\st(\uu \du \sd +~\uu \dd \su + \ud \du \su)
\vac$ \\
          & $\Xi^{*0}$      & $\ss(\su \su \ud +~2 \su \sd \uu) \vac$ \\ \hline
$J = 3/2$ & $\Sigma_c^{*++}$ & $\ss(\uu \uu \cd +~2 \uu \ud \cu) \vac$ \\
($C = 1$) & $\Xi_c^{*+}$    & $\st(\uu \su \cd +~\uu \sd \cu +~\ud \su \cu)
\vac$ \\ \hline
\end{tabular}
\end{center}
\end{table}
\bigskip

\leftline{\bf B.  Representation of pion emission}
\bigskip

Pion emission is represented by a linear combination of products of one
annihilation and one creation operator.  We evaluate the matrix elements of
the following operators between baryon states:
$$
{\cal O}^{(\pi^-)} = u^+ \up d \up - u^+ \dn d \dn~~~,
$$
$$
{\cal O}^{(\pi^0)} = \s(u^+ \up u \up - u^+ \dn u \dn - d^+ \up d \up + d^+ \dn
d \dn)~~~,
$$
\beq
{\cal O}^{(\pi^+)} = -(d^+ \up u \up - d^+ \dn u \dn)~~~.
\eeq
The signs are chosen in accord with standard Clebsch-Gordan coefficient
conventions.
\bigskip

\leftline{\bf C.  Calculation of matrix elements}
\bigskip

We factor matrix elements for specific transitions $A(B\up \to \pi B'\up)$
into isospin Clebsch-Gordan coefficients $(I_B I_{3B}|1 I_{3 \pi} I_{B'}
I_{3B'})$ and isoscalar factors $(\pi B'|B)$:
\beq
A(B\up \to \pi B'\up) = (\pi B'|B)(I_B I_{3B}|1 I_{3 \pi} I_{B'} I_{3B'})~~~.
\eeq
The isoscalar factors are shown in Table II.  The partial widths $\Gamma
(B \to \pi B')$ are just proportional to the squares of these isoscalar
factors, since the squares of Clebsch-Gordan coefficients for respective
charge states sum to unity.  Specifically, we have
\beq
\Gamma(B \to \pi B') = C |(\pi B'|B)|^2 p_{\rm c.m.}^3~~~,
\eeq
where $C$ is a universal constant and the factor $p_{\rm c.m.}^3$ is
appropriate for P-wave decays.  The value of this quantity for each decay is
also shown in Table II, as is the observed partial width.  Unless otherwise
indicated, we quote the best-known partial width \cite{PDG} for a given
isospin multiplet.

\renewcommand{\arraystretch}{1.3}
\begin{table}
\caption{Isoscalar factors $(\pi B'|B)$ for decays of $J = 1/2$ or
$J = 3/2$ baryons to a pion and a $J = 1/2$ baryon.}
\begin{center}
\begin{tabular}{c c c c} \hline
Decay                     & Value of & $p_{\rm c.m.}^3$ & Partial  \\
$B \to \pi B'$           & $(\pi B'|B)$   & (MeV/$c$) & Width (MeV) \\ \hline
$\Delta \to \pi N$       & $2\sqrt{2/3}$ & 227       & $120 \pm 5$ \\
$\Sigma^* \to \pi \Lambda$ & $2/\sqrt{3}$ & 208      & $31.5 \pm 1.0$ \\
$\Sigma^* \to \pi \Sigma$ & $2\sqrt{2}/3$ & 127      & $4.3 \pm 0.7$ \\
$\Xi^* \to \pi \Xi $      & $2/\sqrt{3}$  & 152      & $9.1 \pm 0.5$ \\ \hline
$\Sigma_c \to \pi \Lambda_c$   & $\sqrt{2/3}$ & 91   & \\
$\Sigma^*_c \to \pi \Lambda_c$ & $2/\sqrt{3}$ & (a)  & \\
$\Xi_c^* \to \pi \Xi_c^{(a)}$  & $-1$         & 107  &$< 5.5$ (b) \\
$\Xi_c^* \to \pi \Xi_c^{(s)}$  & $-1/\sqrt{3}$ & (c) & \\ \hline
\end{tabular}
\end{center}
\leftline{(a) (168,~192,~213) MeV/$c$ for $M(\Sigma_c^*) = (2510,~2530,~2550)$
 MeV/$c^2$}
\leftline{(b) 90\% c.l. limit \protect\cite{xicstar}}
\leftline{(c) Unphysical decay}
\end{table}
\bigskip

\centerline{\bf III.  NON-CHARMED BARYON DECAYS}
\bigskip

We may test the relations implied by Table II for SU(3) breaking using the
decays \cite{PDG} of the charmless $J = 3/2$ baryons (the first four rows).
\bigskip

\leftline{\bf A. Prediction for $\Sigma^* \to \pi \Lambda$}
\bigskip

The observed partial width for $\Delta \to \pi N$ implies
\beq
\Gamma(\Sigma^* \to \pi \Lambda)_{\rm pred} = \frac{1}{2} \left(
\frac{208}{227} \right)^3 (120 \pm 5)~\M = 46 \pm 2~\M~~~,
\eeq
The observed value of $31.5 \pm 1.0$ MeV is about $0.68 \pm 0.05$ times the
prediction.
\bigskip

\leftline{\bf B. Prediction for $\Sigma^* \to \pi \Sigma$}
\bigskip

The observed partial width for $\Sigma^* \to \pi \Lambda$ implies
\beq
\Gamma(\Sigma^* \to \pi \Sigma)_{\rm pred} = \frac{2}{3} \left( \frac{127}{208}
\right)^3 (31.5 \pm 1.0)~\M = 4.8 \pm 0.2~\M~~~,
\eeq
The observed value of $4.3 \pm 0.7$ MeV is in satisfactory agreement with the
prediction.
\bigskip

\leftline{\bf C.  Prediction for $\Xi^* \to \pi \Xi$}
\bigskip

The observed partial width for $\Sigma^* \to \pi \Lambda$ implies
\beq
\Gamma(\Xi^* \to \pi \Xi)_{\rm pred} = \left( \frac{152}{208}
\right)^3 (31.5 \pm 1.0)~\M = 12.3 \pm 0.4~\M~~~,
\eeq
The observed value of $9.1 \pm 0.5$ MeV is about $0.74 \pm 0.05$ times the
prediction.
\bigskip

\leftline{\bf D.  Systematics of SU(3) breaking}
\bigskip

It appears that the replacement of a nonstrange by a strange quark multiplies
the decay width by a factor of approximately 0.7.  We will bear this factor in
mind when discussing possible violations of the symmetry which involves
replacing a strange quark by a charmed quark.  First-order symmetry breaking in
the above decays has been discussed, for example, in Ref.~\cite{Butler}.
\newpage

\centerline{\bf IV.  CHARMED BARYON DECAYS}
\bigskip

\leftline{\bf A. $\Sigma_c \to \pi \Lambda_c$}
\bigskip

This decay is kinematically allowed, in contrast to the decay $\Sigma \to \pi
\Lambda$.  The small c.m. momentum leads to a small predicted width:
\beq
\Gamma(\Sigma_c \to \pi \Lambda_c)  = \frac{1}{2} \left( \frac{91}
{208} \right)^3 (31.5 \pm 1.0)~\M = 1.32 \pm 0.04~\M~~~,
\eeq
narrower than the experimental resolution with which this state is seen.  Here
we have related $\Sigma_c \to \pi \Lambda_c$ to $\Sigma^* \to \pi \Lambda$;
both processes involve baryons with two nonstrange quarks.
\bigskip

\leftline{\bf B. $\Sigma_c^* \to \pi \Lambda_c$}
\bigskip

This decay is the analogue of $\Sigma^* \to \pi \Lambda$ under the replacement
$s \to c$.  The isoscalar factors are the same for the two decays, so the
ratio of partial widths in the limit of exact symmetry under $s \leftrightarrow
c$ should simply scale according to the ratio of the values of $p_{\rm cm}^3$.

We may estimate the mass of $\Sigma_c^*$ by means of a simple hyperfine
splitting calculation \cite{DGG,Sakh,GR}.  In the $\Sigma^{(*)+}$ and
$\Sigma_c^{(*)++}$ states, the two $u$ quarks are coupled to a total spin of
1, so that one expects the splitting between the $J = 1/2$ and $J = 3/2$
baryonic states to scale as $1/m_Q$, where $Q = s$ or $c$.  Thus if the wave
function of the $J = 1$ diquark is the same at the positions of the $s$ quark
in the $\Sigma^{(*)}$ and the $c$ quark in the $\Sigma_c^{(*)}$, we expect
\beq \label{eqn:sig}
M(\Sigma_c^*) = M(\Sigma_c) + (m_s/m_c) [ M(\Sigma^*) - M(\Sigma) ]
\approx 2514~\M~~~,
\eeq
where we have used constituent-quark masses \cite{GR} $m_s = 538$ MeV/$c^2$
and $m_c = m_s + M(\Lambda_c) - M(\Lambda) = 1707$ MeV/$c^2$.  A similar
attempt to relate the hyperfine splitting $M(D_s^*)-M(D_s)$ to $M(D^*)-M(D)$
underestimates the former \cite{RWi}; reduced-mass effects apparently cannot
be ignored. Similarly, we expect that the hyperfine splitting between
$\Sigma_c^*$ and $\Sigma_c$ will, if anything, exceed the na\"{\i}ve estimate.
Hence (\ref{eqn:sig}) should be regarded as a lower bound.  Other theoretical
estimates for charmed baryon masses (see, e.g., \cite{Sakh}, \cite{GR},
\cite{theory}, and \cite{fest}) lead one to expect $M(\Sigma_c^*)$ between
about 2.50 and 2.55 GeV/$c^2$.

In Fig.~2 we have plotted the total width of $(\Sigma_c^*)$, approximately
equal to the partial width for $\Sigma_c^* \to \pi \Lambda_c$ aside from small
electromagnetic transitions, as a function of $M(\Sigma^*_c)$.  Since these
predictions were obtained from $\Gamma(\Sigma^* \to \pi \Lambda)$ via the
substitution $s \to c$, and we have seen that substituting heavier spectator
quarks reduces partial widths in the case of non-charmed baryons, it is
reasonable to expect the predictions of Fig.~2 to be upper bounds.  These
widths probably exceed available mass resolutions in CLEO or various
fixed-target Fermilab experiments, so that optimum signal-to-noise advantages
with respect to combinatorial backgrounds are not being achieved in the search
for a $\Sigma_c^*$. Nevertheless, the widths in Fig.~2 are sufficiently modest
that it should not be too hard to find this state.
\bigskip

\begin{figure}
\centerline{\epsfysize = 4in \epsffile{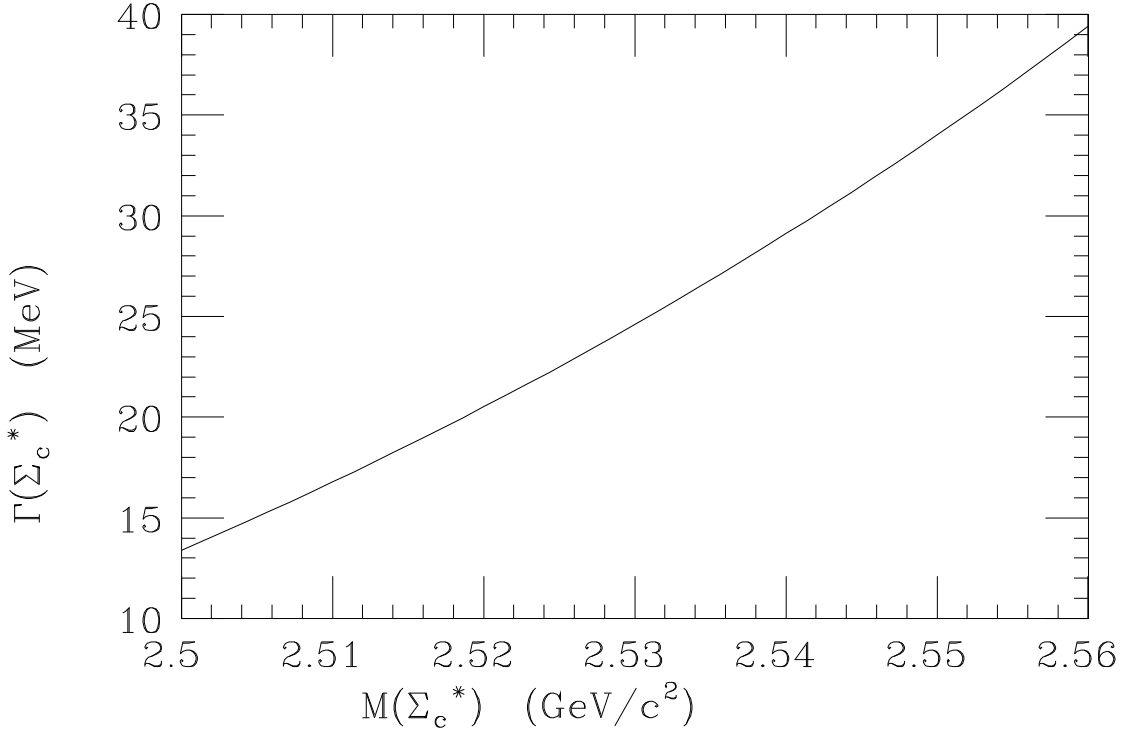}}
\caption{Predicted total width $\Gamma(\Sigma_c^*) \approx \Gamma(\Sigma_c^*
\to \pi \Lambda_c)$ as a function of the mass of $\Sigma_c^*$.}
\end{figure}

\leftline{\bf C.  Decays of the $\Xi_c^*$}
\bigskip

The state discovered by CLEO \cite{xicstar} at a mass of $2643$ MeV/$c^2$ lies
30 MeV/$c^2$ above a na\"{\i}ve estimate \cite{fest} which was based on
assuming a universal hyperfine interaction proportional to the inverse of
products of quark masses.  Thus, the $\Xi_c^* - \Xi_c^{(s)}$ splitting appears
to be about $178 - 95 = 83$ MeV/$c^2$ instead of the 53 MeV/$c^2$ estimated in
Ref.~\cite{fest}.  Nonetheless, the phase space for the decay $\Xi_c^* \to \pi
\Xi_c^{(a)}$ remains small enough that we predict a small partial width.  The
process $\Xi^* \to \pi \Xi$ involves two strange-quark spectators, whereas the
spectators in $\Xi_c^* \to \pi \Xi_c^{(a)}$ are one strange and one charmed
quark.  Thus we expect $\Xi^* \to \pi \Xi$ to provide the best reference
amplitude; if anything, the partial width for $\Xi_c^* \to \pi \Xi_c^{(a)}$
will be no larger than the following prediction:
\beq
\Gamma(\Xi_c^* \to \pi \Xi_c^{(a)}) = \frac{3}{4} \left( \frac{106}{152}
\right)^3 (9.1 \pm 0.5)~\M = 2.3 \pm 0.1~\M~~~.
\eeq

The decay $\Xi_c^* \to \pi \Xi_c^{(s)}$ is kinematically forbidden.  The square
of its isoscalar factor is only 1/3 of that for the allowed decay $\Xi_c^* \to
\pi \Xi_c^{(a)}$.
\bigskip

\leftline{\bf D.  Relations between hyperfine splittings}
\bigskip

An elementary calculation along the lines of Ref.~\cite{GR} leads to the
relation
\beq
\frac{M(\Xi_c^*) - M(\Xi_c^{(s)})}{M(\Sigma_c^*) - M(\Sigma_c)} = \frac{1}{2}
\left( 1 + \frac{m_u}{m_s} \right)~~~
\eeq
in the limit of universal hyperfine interactions mentioned earlier.  Given
the likelihood that the $\Xi_c^{(*,s)}$ wave functions are spatially more
compact than those of the $\Sigma_c^{(*)}$ states, this relation must be
regarded as a {\it lower bound}, implying an upper bound on $M(\Sigma_c^*) -
M(\Sigma_c)$ .  Taking \cite{GR} $m_u = 363$ MeV/$c^2$ and $m_s = 538$
MeV/$c^2$, we find $M(\Sigma_c^*) - M(\Sigma_c) \le 99$ MeV/$c^2$, or
$M(\Sigma_c^*) \le 2552$ MeV/$c^2$.  Referring to Fig.~2, we see that the width
of this state should not exceed 35 MeV.

One can perform a similar calculation to estimate the hyperfine splitting
between $\Omega_c^*$ and $\Omega_c$.  We find
\beq
\frac{M(\Omega_c^*) - M(\Omega_c)}{M(\Xi_c^*) - M(\Xi_c^{(s)})} = \frac{2}
{1 + \frac{m_s}{m_u}}~~~,
\eeq
leading to the prediction $M(\Omega_c^*) - M(\Omega_c) = 67$ MeV/$c^2$, or
$M(\Omega_c^*) = 2771$ MeV/$c^2$.  For the same reasons as mentioned above,
we expect symmetry-breaking in the wave function to increase the hyperfine
splitting and the $\Omega_c^*$ mass.  Thus the figure we quote is a lower
limit.  The decay $\Omega_c^* \to \Omega_c \gamma$ will be the means for
detecting the $\Omega_c^*$.

An equal-spacing rule follows from the assumptions \cite{Savage}
of heavy-quark symmetry and lowest-order SU(3) symmetry breaking.  In this
approach the corrections due to chiral loops are found to be finite and
small.  We can also obtain such a rule by linearizing our hyperfine expressions
in $m_s - m_d$.  One obtains
$$
M(\Xi_c^{(s)}) - M(\Sigma_c) = M(\Omega_c) - M(\Xi_c^{(s)})
$$
\beq
= M(\Xi_c^*) - M(\Sigma_c^*) = M(\Omega_c^*) - M(\Xi_c^*)~~~.
\eeq
The relations between states with a given $J = 1/2$ or $3/2$ follow from the
fact that the product ${\bf 6} \times {\bf 6^*} = {\bf 1} + {\bf 8} + {\bf
27}$ contains a single octet, but the relations between states with $J = 1/2$
and states with $J = 3/2$ are the consequence of the heavy-quark symmetry.
Experimentally $M(\Xi_c^{(s)})
- M(\Sigma_c) \approx 110~\M/c^2$, while $M(\Omega_c) - M(\Xi_c^{(s)}) \approx
141~\M/c^2$.  We have predicted $91~\M/c^2 \le M(\Xi_c^*) - M(\Sigma_c^*) \le
129~\M/c^2$ and $125~\M/c^2 \le M(\Omega_c^*) - M(\Xi_c^*)$.
\newpage

\leftline{\bf E.  Production of the $\Sigma_c^*$}
\bigskip

The failure of the $\Sigma_c^*$ to be produced abundantly (in contrast to the
$\Sigma^*$ discussed in Sec.~III) may be due in part to the difference between
mass splittings in the strange and charmed sectors.  The $\Sigma^*$ can just
barely be produced via the S-wave decay of the lowest-lying spin-3/2 excited
$\Lambda$ state, $\Lambda(1520) \to \pi \Sigma^*(1385)$.  In contrast, a
candidate for the lowest-lying spin-3/2 excited $\Lambda_c$ at 2626 MeV/$c^2$
\cite{PDG,fest} lies too low in mass to decay to $\pi \Sigma_c^*(> 2500)$.
\bigskip

\centerline{\bf V.  CONCLUSIONS}
\bigskip

We have discussed the pionic decays of non-charmed and charmed baryons in an
attempt to understand the small width of the recently observed \cite{xicstar}
candidate for the spin-3/2 state $\Xi_c^*$.  Relating the decay $\Xi_c^* \to
\pi \Xi_c^{(a)}$ to the process $\Xi^* \to \pi \Xi$, we predict $\Gamma(\Xi_c^*
\to \pi \Xi_c^{(a)}) = 2.3$ MeV in the limit in which strange and charmed
spectator quarks are interchangeable.  In fact, this prediction is more likely
to be an upper bound.

We have shown that the $\Sigma_c^*$, so far claimed in only one experiment
\cite{Ammosov}, should have a total width modestly exceeding the mass
resolution of most present-day experiments but not more than 35 MeV, and a
mass not exceeding 2552 MeV/$c^2$.  Evidence for this state (or confirmation
of the results of Ref.~\cite{Ammosov}) and a reliable width measurement would
permit the recalibration of pionic decay widths of charmed baryons, for which
present predictions rely on an extrapolation from the charmless sector.  The
detection of a state $\Omega_c^*$ with a mass of at least 2771 MeV/$c^2$
would then complete the picture of the singly-charmed ground-state baryons.
\bigskip

\centerline{\bf ACKNOWLEDGMENTS}
\bigskip

I am indebted to D. O. Riska, M. Savage and J. Yelton for useful discussions.
I thank the Institute for Nuclear Theory at the University of Washington for
hospitality during this work, which was supported in part by the United States
Department of Energy under Grant No. DE FG02 90ER40560.
\bigskip

\def \ajp#1#2#3{Am. J. Phys. {\bf#1}, #2 (#3)}
\def \apny#1#2#3{Ann. Phys. (N.Y.) {\bf#1}, #2 (#3)}
\def \app#1#2#3{Acta Phys. Polonica {\bf#1}, #2 (#3)}
\def \arnps#1#2#3{Ann. Rev. Nucl. Part. Sci. {\bf#1}, #2 (#3)}
\def \cmts#1#2#3{Comments on Nucl. Part. Phys. {\bf#1}, #2 (#3)}
\def \cn{Collaboration}
\def \cp89{{\it CP Violation,} edited by C. Jarlskog (World Scientific,
Singapore, 1989)}
\def \efi{Enrico Fermi Institute Report No. EFI}
\def \f79{{\it Proceedings of the 1979 International Symposium on Lepton and
Photon Interactions at High Energies,} Fermilab, August 23-29, 1979, ed. by
T. B. W. Kirk and H. D. I. Abarbanel (Fermi National Accelerator Laboratory,
Batavia, IL, 1979}
\def \hb87{{\it Proceeding of the 1987 International Symposium on Lepton and
Photon Interactions at High Energies,} Hamburg, 1987, ed. by W. Bartel
and R. R\"uckl (Nucl. Phys. B, Proc. Suppl., vol. 3) (North-Holland,
Amsterdam, 1988)}
\def \ib{{\it ibid.}~}
\def \ibj#1#2#3{~{\bf#1}, #2 (#3)}
\def \ichep72{{\it Proceedings of the XVI International Conference on High
Energy Physics}, Chicago and Batavia, Illinois, Sept. 6 -- 13, 1972,
edited by J. D. Jackson, A. Roberts, and R. Donaldson (Fermilab, Batavia,
IL, 1972)}
\def \ijmpa#1#2#3{Int. J. Mod. Phys. A {\bf#1}, #2 (#3)}
\def \ite{{\it et al.}}
\def \jpb#1#2#3{J.~Phys.~B~{\bf#1}, #2 (#3)}
\def \lkl87{{\it Selected Topics in Electroweak Interactions} (Proceedings of
the Second Lake Louise Institute on New Frontiers in Particle Physics, 15 --
21 February, 1987), edited by J. M. Cameron \ite~(World Scientific, Singapore,
1987)}
\def \ky85{{\it Proceedings of the International Symposium on Lepton and
Photon Interactions at High Energy,} Kyoto, Aug.~19-24, 1985, edited by M.
Konuma and K. Takahashi (Kyoto Univ., Kyoto, 1985)}
\def \mpla#1#2#3{Mod. Phys. Lett. A {\bf#1}, #2 (#3)}
\def \nc#1#2#3{Nuovo Cim. {\bf#1}, #2 (#3)}
\def \np#1#2#3{Nucl. Phys. {\bf#1}, #2 (#3)}
\def \PDG{Particle Data Group, L. Montanet \ite, \prd{50}{1174}{1994}}
\def \pisma#1#2#3#4{Pis'ma Zh. Eksp. Teor. Fiz. {\bf#1}, #2 (#3) [JETP Lett.
{\bf#1}, #4 (#3)]}
\def \pl#1#2#3{Phys. Lett. {\bf#1}, #2 (#3)}
\def \pla#1#2#3{Phys. Lett. A {\bf#1}, #2 (#3)}
\def \plb#1#2#3{Phys. Lett. B {\bf#1}, #2 (#3)}
\def \pr#1#2#3{Phys. Rev. {\bf#1}, #2 (#3)}
\def \prc#1#2#3{Phys. Rev. C {\bf#1}, #2 (#3)}
\def \prd#1#2#3{Phys. Rev. D {\bf#1}, #2 (#3)}
\def \prl#1#2#3{Phys. Rev. Lett. {\bf#1}, #2 (#3)}
\def \prp#1#2#3{Phys. Rep. {\bf#1}, #2 (#3)}
\def \ptp#1#2#3{Prog. Theor. Phys. {\bf#1}, #2 (#3)}
\def \rmp#1#2#3{Rev. Mod. Phys. {\bf#1}, #2 (#3)}
\def \rp#1{~~~~~\ldots\ldots{\rm rp~}{#1}~~~~~}
\def \si90{25th International Conference on High Energy Physics, Singapore,
Aug. 2-8, 1990}
\def \slc87{{\it Proceedings of the Salt Lake City Meeting} (Division of
Particles and Fields, American Physical Society, Salt Lake City, Utah, 1987),
ed. by C. DeTar and J. S. Ball (World Scientific, Singapore, 1987)}
\def \slac89{{\it Proceedings of the XIVth International Symposium on
Lepton and Photon Interactions,} Stanford, California, 1989, edited by M.
Riordan (World Scientific, Singapore, 1990)}
\def \smass82{{\it Proceedings of the 1982 DPF Summer Study on Elementary
Particle Physics and Future Facilities}, Snowmass, Colorado, edited by R.
Donaldson, R. Gustafson, and F. Paige (World Scientific, Singapore, 1982)}
\def \smass90{{\it Research Directions for the Decade} (Proceedings of the
1990 Summer Study on High Energy Physics, June 25--July 13, Snowmass,
Colorado),
edited by E. L. Berger (World Scientific, Singapore, 1992)}
\def \tasi90{{\it Testing the Standard Model} (Proceedings of the 1990
Theoretical Advanced Study Institute in Elementary Particle Physics, Boulder,
Colorado, 3--27 June, 1990), edited by M. Cveti\v{c} and P. Langacker
(World Scientific, Singapore, 1991)}
\def \yaf#1#2#3#4{Yad. Fiz. {\bf#1}, #2 (#3) [Sov. J. Nucl. Phys. {\bf #1},
#4 (#3)]}
\def \zhetf#1#2#3#4#5#6{Zh. Eksp. Teor. Fiz. {\bf #1}, #2 (#3) [Sov. Phys. -
JETP {\bf #4}, #5 (#6)]}
\def \zpc#1#2#3{Zeit. Phys. C {\bf#1}, #2 (#3)}
\def \zpd#1#2#3{Zeit. Phys. D {\bf#1}, #2 (#3)}

\end{document}